\documentclass{article}
\usepackage[T1]{fontenc}
\usepackage[utf8]{inputenc}
\usepackage{ismir} 
\usepackage{amsmath,cite,url}
\usepackage{graphicx}
\usepackage{color}
\usepackage{multirow, amssymb} 

\title{Estimating the Reliability of Dynamic Time Warping Alignments Using Circumstantial Evidence}

\oneauthor
{Aanya Pratapneni \qquad Alice Yuan \qquad TJ Tsai}
{\\Harvey Mudd College}

\def\authorname{A. Pratapneni, A. Yuan, and T. Tsai}

\begin{document}

\maketitle

\begin{abstract}
Recent works have explored ways to handle uncertainty in dynamic time warping (DTW) alignment paths through the use of differentiable variants of DTW like Soft-DTW. In this paper, we approach the issue of uncertainty in DTW alignment paths in a different way. Given a DTW alignment path, we propose a metric that indicates how reliable a local segment of the alignment path is. The intuition for our metric is based on the idea of circumstantial evidence. If DTW has found a very prominent path, then if we re-run the alignment with relaxed boundary conditions, it will still pick the same path. If, on the other hand, DTW has found a “weak” path, then re-running the alignment with relaxed boundary conditions will likely yield a different path. Accordingly, our reliability metric is computed by picking a local section of the DTW alignment path, re-estimating the alignment with FlexDTW (which allows flexibility in the boundary conditions), and then measuring how well the DTW and FlexDTW paths agree. We assess the proposed reliability metric on DTW alignment paths containing both matching and non-matching regions across a range of scenarios on an audio-audio alignment task. We find that the reliability metric correctly identifies reliable regions of the alignment path with an aggregate AUROC of $0.97$. This approach provides an unsupervised method for estimating the reliability of a DTW alignment path.
\end{abstract}

\section{Introduction}
\label{sec:introduction}

Dynamic Time Warping (DTW) is a dynamic programming algorithm for finding the optimal alignment between two sequences. It is widely used in the music information retrieval literature for offline alignment of two unstructured music sequences \cite{muller2021fundamentals}. Given a set of allowable transitions and corresponding weights, DTW finds a path through the pairwise cost matrix that has the lowest cumulative path score. However, it does not provide an indication of how reliable the estimated alignment path is. The left part of Figure 1 shows an example where part of a DTW alignment path is inaccurate because the two performances use different cadenzas (the start and end of the cadenza section is indicated by dotted lines). In this paper, we explore a way to estimate the reliability of DTW alignment paths based on circumstantial evidence. The right part of Figure 1 shows a visualization of our proposed reliability metric, which indicates the non-matching region of the alignment path in red.

\begin{figure}
 \centering
 \includegraphics[alt={A DTW alignment where part of the alignment path is unreliable},width=\linewidth]{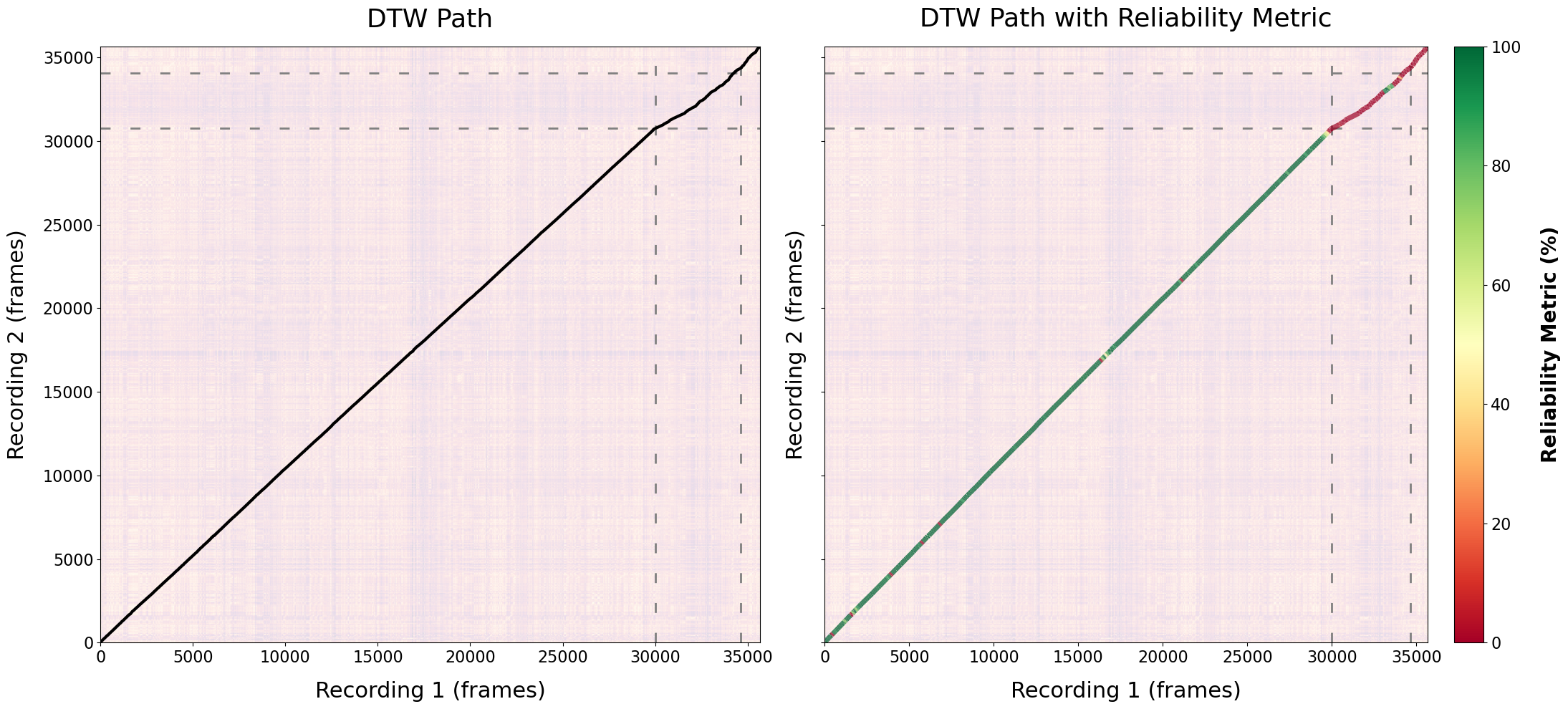}
 \caption{An example where the estimated DTW alignment path between two recordings is unreliable in a certain region (indicated by dotted lines) due to differing cadenzas.  Our proposed reliability metric indicates how reliable the predicted DTW alignment path is in different regions.}
 \label{fig:motivation}
\end{figure}

Many previous works on DTW in the music and audio research community fall into two groups. The first group focuses on mitigating the quadratic runtime and memory requirements of DTW. Some examples in the audio and music literature include imposing fixed global constraint bands in the cost matrix \cite{itakura1975minimum, sakoe1978dynamic}, estimating the alignment at multiple resolutions \cite{PraetzlichDM16_MsDTW_ICASSP} and/or working with a fixed amount of memory by imposing local constraint regions \cite{PraetzlichDM16_MsDTW_ICASSP}, and exploring parallelizable approximations of DTW \cite{tsai2021segmental}. Tralie and Dempsey \cite{tralie2020exact} compute exact DTW with linear memory by processing the cost matrix across diagonals rather than rows and columns. The broader data mining community has also explored ways to speed up computation through specialized hardware \cite{sart2010accelerating, wang2013accelerating}, multi-resolution approaches \cite{SalvadorC04_fastDTW}, using lower bounds to reduce computation \cite{rakthanmanon2012searching, zhang2011inner}, and approximations of DTW distance \cite{lods2017learning, nagendar2015efficient}. The second group focuses on modifying the behavior of DTW to make it more flexible. Some examples in the music information retrieval literature include aligning sequences in an online fashion \cite{dixon2005live, macrae2010accurate}, handling structural differences like jumps and repeats \cite{shan2020improved, Grachten2013AutomaticAO}, allowing for flexible boundary conditions \cite{bukey2023flexdtw}, doing partial alignments \cite{MuellerA08_PathConstrained_ICASSP}, being robust to key drift in a capella recordings \cite{waloschek2018driftin}, and using multiple performances to improve alignment accuracy \cite{wang2016robust}.

In recent years, a line of research has explored ways to handle uncertainty in DTW alignment paths. Soft-DTW \cite{cuturi2017soft, mensch2018differentiable} offers a differentiable variant of DTW in which the hard minimum operator is replaced with a smooth approximation. Soft-DTW has been explored in the MIR literature on topics like performance-score synchronization \cite{agrawal2021convolutional} and multi-pitch estimation \cite{krause2023soft, krause2023weakly}, and subsequent works have proposed improved practices to stabilize training \cite{ZeitlerDKM23_StablizeSoftDTW_ISMIR}, use customizable step weights \cite{zeitler2024soft}, and achieve desirable mathematical properties \cite{blondel2021differentiable}. Recently, Zeitler and Muller \cite{zeitler2026unified} proposed a framework called Differentiable DTW (dDTW) that includes Soft-DTW and CTC loss as special cases.  A few works have addressed alignment confidence more directly: Raffel and Ellis \cite{raffel2015large, raffel2016optimizing} use a normalized path cost as a confidence score for the entire DTW alignment path, and Maman and Bermano \cite{maman2022unaligned} use heuristics based on ``singular points'' to discard unrealiable alignment fragments.  While this work focuses on DTW, probabilistic models such as hidden Markov models more naturally support uncertainty quantification, e.g., through the posterior variance of a tracked position \cite{maezawa2017muens}.

This paper explores the issue of uncertainty in DTW alignment paths, but in a different way. Given a DTW alignment path, we focus on developing a metric that indicates how reliable the predicted alignment path is. The intuition for our metric is simple and based on the idea of circumstantial evidence. If DTW has found a very ``strong'' path, then if we re-run the alignment with relaxed boundary conditions, it will still pick the same path. If, on the other hand, DTW has found a “weak” path (e.g., the best path through noise), then if we re-run the alignment with relaxed boundary conditions, it will likely pick a different path. Accordingly, our method picks a section of the DTW alignment path, re-runs the alignment but gives the alignment path much more freedom in its boundary conditions, and then observes how much the alignment path changes. If the alignment path does not change at all, this provides a strong indicator that DTW has found a prominent alignment path. This method provides a way to identify reliable alignments in an unsupervised manner.  

This paper has two main contributions. First, we introduce an unsupervised method for estimating the reliability of a DTW alignment path. The method consists of taking small sections of the DTW alignment path, re-running the alignment on the corresponding local chunk of the pairwise cost matrix with FlexDTW (which allows for flexible boundary conditions), and measuring how well the DTW and FlexDTW alignment paths agree. High agreement provides strong circumstantial evidence that DTW has found a “strong” path, and that the alignment path is reliable. Second, we present empirical results that measure how well our reliability metric can detect reliable and unreliable regions of DTW paths. Towards this end, we construct a suite of 19 benchmarks in which the alignment paths have both matching and non-matching regions across a variety of scenarios. We find that our proposed reliability metric is able to discriminate between reliable and unreliable regions, achieving an aggregate area under the receiver operating characteristic curve (AUROC) of $0.970$. We also present analyses to understand the failure modes and limitations of the proposed reliability metric.\footnote{Code for this project can be found at \url{https://github.com/HMC-MIR/DTW-Reliability-Metric}.}


\section{Proposed Method}
\label{sec:method}

\begin{figure}
 \centering
 \includegraphics[alt={Overview of 4 steps in our proposed method},width=\linewidth]{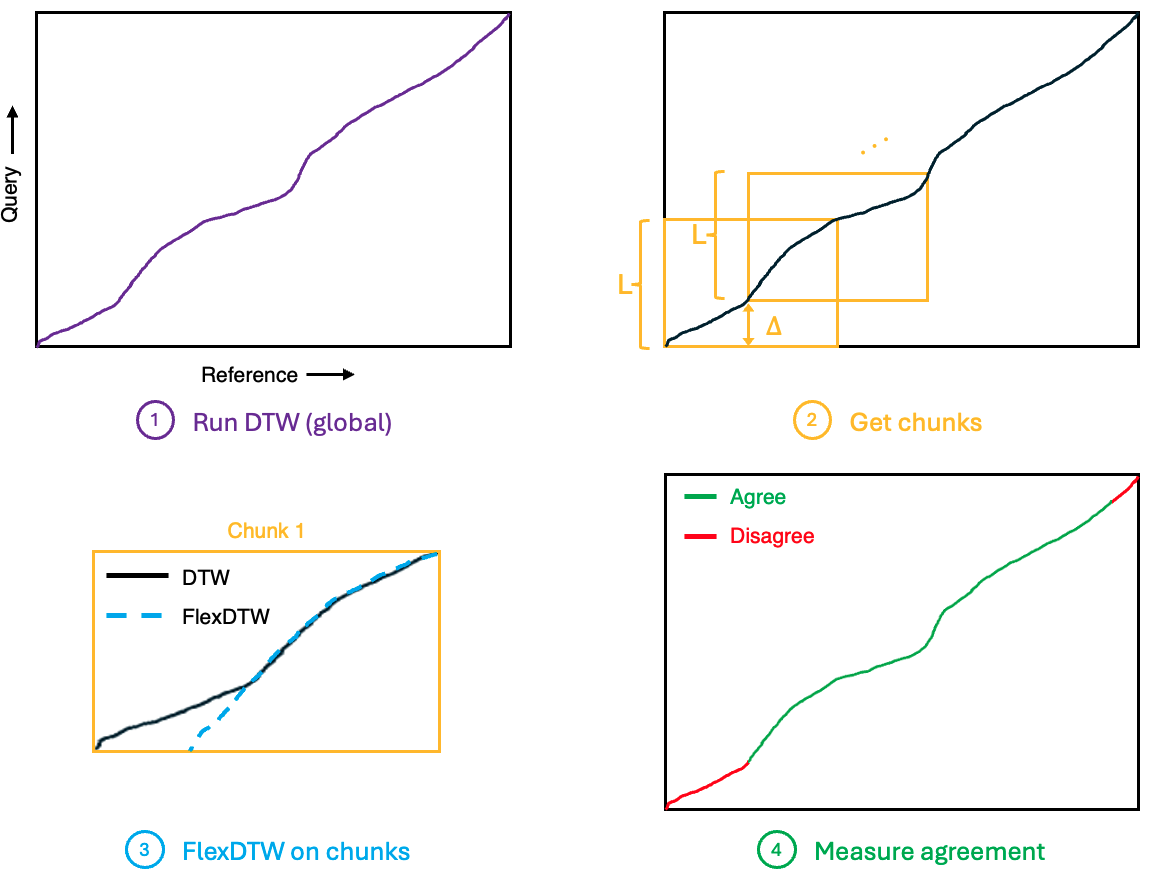}
 \caption{Overview of our proposed method for estimating the reliability of DTW alignments.}
 \label{fig:methodOverview}
\end{figure}

In this section, we describe our proposed method for estimating the reliability of DTW alignments. Figure~\ref{fig:methodOverview} shows an overview of the proposed method, which consists of four main steps. These steps are described in the next four paragraphs.

The first step is to perform DTW on the two sequences. Given two sequences $x_0, x_1, …, x_{N-1}$ and $y_0, y_1, …, y_{M-1}$ that we would like to align, we perform DTW to estimate the optimal alignment between the two sequences. This is accomplished by first computing a pairwise cost matrix $C$ in $R^{N\times M}$ where $C[i,j]$ indicates the dissimilarity between $x_i$ and $y_j$, and then using dynamic programming to find a path through $C$ that has the lowest cumulative cost. The estimated alignment is specified by a sequence of coordinates $(i_0, j_0)$, $(i_1, j_1)$,..., $(i_{T-1}, j_{T-1})$, where $T$ indicates the number of elements in the warping path. Note that in standard DTW, it is assumed a priori that the two sequences begin and end together, so that $(i_0, j_0) = (0,0)$ and $(i_{T-1}, j_{T-1}) = (N-1, M-1)$. If we draw the cost matrix with sequences increasing from left to right and bottom to top (as in Figure~\ref{fig:methodOverview}), then DTW assumes that the alignment path will begin in the lower left corner and end in the upper right corner. The goal of our proposed method is to calculate a reliability score for each element in the DTW warping path.

The second step is to extract a sequence of chunks from the global cost matrix $C$. This is done in the following manner. We first extract a sequence of overlapping windows of length $L$ from the first sequence $x_0, …, x_{N-1}$, similar to the way that analysis windows are formed when computing a Short-Time Fourier Transform. For example, the first window is $(x_0, x_1, …, x_{L-1})$ and the second window is $(x_{\Delta}, x_{\Delta+1}, …, x_{\Delta+L-1})$, where $\Delta$ is the hop size between adjacent windows. For each window of length $L$, we infer the rectangular chunk of the global cost matrix $C$ induced by the window and the estimated DTW warping path. For example, if the window is $(x_{50}, x_{51}, …, x_{99})$ and the DTW warping path aligns $x_{50}$ and $x_{99}$ to $y_{60}$ and $y_{119}$, respectively, then the induced chunk of the cost matrix is $C[50:100, 60:120]$ (expressed in numpy notation). We denote these local chunks of the pairwise cost matrix as $\tilde{C}_i$, where $\tilde{C}$ indicates a local chunk and $i$ indicates the chunk index (e.g., $\tilde{C}_0$ corresponds to the first chunk). Note that the dimensions of $\tilde{C}_i$ are $L \times \tilde{L}_i$, where $L$ is fixed but $\tilde{L}_i$ varies from chunk to chunk, depending on the nature of the DTW warping path. The top right part of Figure 2 shows the first two chunks extracted from the example warping path.

The third step is to perform FlexDTW on each chunk. FlexDTW \cite{bukey2023flexdtw} is a variant of DTW that allows for flexible boundary conditions. If we draw a cost matrix such that sequences increase from left to right and bottom to top (as in Figure~\ref{fig:methodOverview}), then FlexDTW allows alignment paths to begin anywhere on the left or bottom edge, and to end anywhere on the top or right edge. To compare paths of different lengths, FlexDTW selects the “best” path based on a normalized path cost, in which the cumulative path cost is normalized by the path’s Manhattan distance. Importantly, when running FlexDTW on a local chunk, we set the allowable transitions and transition weights to match that of the global DTW, so that the only difference between the two is the flexibility in boundary conditions. At the end of this step, we have two estimated warping paths in each chunk: one from the global DTW warping path, and one from the local FlexDTW warping path. The bottom left part of Figure 2 shows both of these warping paths through $\tilde{C}_0$.

The fourth step is to measure the agreement between the global DTW and local FlexDTW warping paths.  For each local chunk, we compute a single number that describes the percentage of DTW warping path coordinates (only in the local chunk) that ``agree'' with the FlexDTW warping path.  Here, we define ``agreement'' to mean that the DTW coordinate is within some Euclidean distance $\varepsilon$ of any element in the FlexDTW warping path, where $\varepsilon$ is a tolerance that can be set as a hyperparameter.  Note that an appropriate setting of $\varepsilon$ depends on the application and the desired level of precision: if very high precision is required, then $\varepsilon$ should be set to a smaller number so that the reliability metric will only be ``high’’ when the two paths are very close to each other.  In our experiments, we used $\varepsilon = 10$ frames, which corresponds to about 232ms.  Thus, ``reliable'' means trustworthy at the chosen tolerance $\varepsilon$, rather than correct per se.  Once we have computed an agreement metric on a local chunk, we impute that agreement value to every DTW coordinate in that local chunk.  Since the chunks are overlapping, most DTW coordinates will have multiple agreement metrics imputed to it.  We calculate an aggregate metric of agreement as the median of any imputed agreement values.  So, for example, if $L=300$ and $\Delta = 100$, then most DTW coordinates would be included in three chunks, and the aggregate metric of agreement would be the median of the three values.  At the end of this fourth step, every coordinate on the global DTW warping path has an agreement metric that indicates how well it agrees with its surrounding local FlexDTW warping paths.  The bottom right part of Figure~\ref{fig:methodOverview} shows how an agreement metric can indicate regions of the DTW warping path that may not be reliable.

\section{Experimental Setup}
\label{sec:expSetup}

In this section, we describe an experimental setup that allows us to assess the utility of our proposed reliability metric.

\textit{Approach}.  The proposed reliability metric is an indicator of how much we can trust a DTW warping path.  To assess the utility of this reliability metric, we construct an alignment benchmark in which the alignment paths contain both matching and non-matching regions.  Our approach is to start with an existing benchmark for measuring audio-audio alignment accuracy, and then modify the data in a controlled manner by inserting non-matching data into one of the sequences.  We can then measure how well the reliability metric identifies parts of the alignment path as reliable or unreliable.  To study the effects of location and duration of non-matching regions, we construct a set of benchmarks that span a range of scenarios.

\textit{Original benchmark}.  The original source data is the Chopin Mazurka dataset \cite{sapp2008hybrid}. This dataset contains historic recordings of five different Chopin Mazurkas, where each recording comes with manually annotated beat-level timestamps. This dataset has been used in previous studies to measure the accuracy of audio-audio alignment techniques \cite{sapp2008hybrid, grosche2010makes, schreiber2020modeling}. Table 1 contains an overview of the original dataset. 

\begin{table}
\begin{center}
	\begin{tabular}{|l|c|c|c|c|c|} 
		\hline
		Piece & Files & mean & std & min & max \\
		\hline
		Opus 17, No 4 & 64 & 259.7 & 32.5 & 194.4 & 409.6 \\
		Opus 24, No 2 & 64 & 137.5 & 13.9 & 109.6 & 180.0 \\
		Opus 30, No 2 & 34 & 85.0 & 9.2 & 68.0 & 99.0 \\
		Opus 63, No 3 & 88 & 129.0 & 13.4 & 96.2 & 162.9 \\
		Opus 68, No 3 & 51 & 101.1 & 19.4 & 71.8 & 164.8 \\
		\hline
	\end{tabular}
\end{center}
\caption{Overview of the Chopin Mazurka dataset.  All durations are in seconds.  These recordings were modified in a controlled manner to assess how well our reliability metric can discriminate between reliable and unreliable parts of the DTW alignment path.}
\label{tab:data}
\end{table}

\textit{DTW Settings}.  We calculate the proposed reliability metric on a DTW warping path with allowed transitions $(1,1), (1,2), (2,1)$ and corresponding weights $2, 3, 3$. To construct the pairwise cost matrix between two recordings, we used librosa’s chroma\_cqt function and a cosine distance metric.

\textit{Modified Benchmarks:~Overview}.  We modify the Mazurka dataset to construct scenarios in which the DTW warping path has both matching and non-matching regions.  We can then study how well our reliability metric can detect incorrect or unreliable parts of the warping path.  We construct a total of 19 different benchmarks, each focusing on a different scenario.

\begin{figure}
 \centering
 \includegraphics[alt={Four different scenarios in the benchmark suite: fully matching, fully non-matching, mostly matching, and mostly non-matching},width=\linewidth]{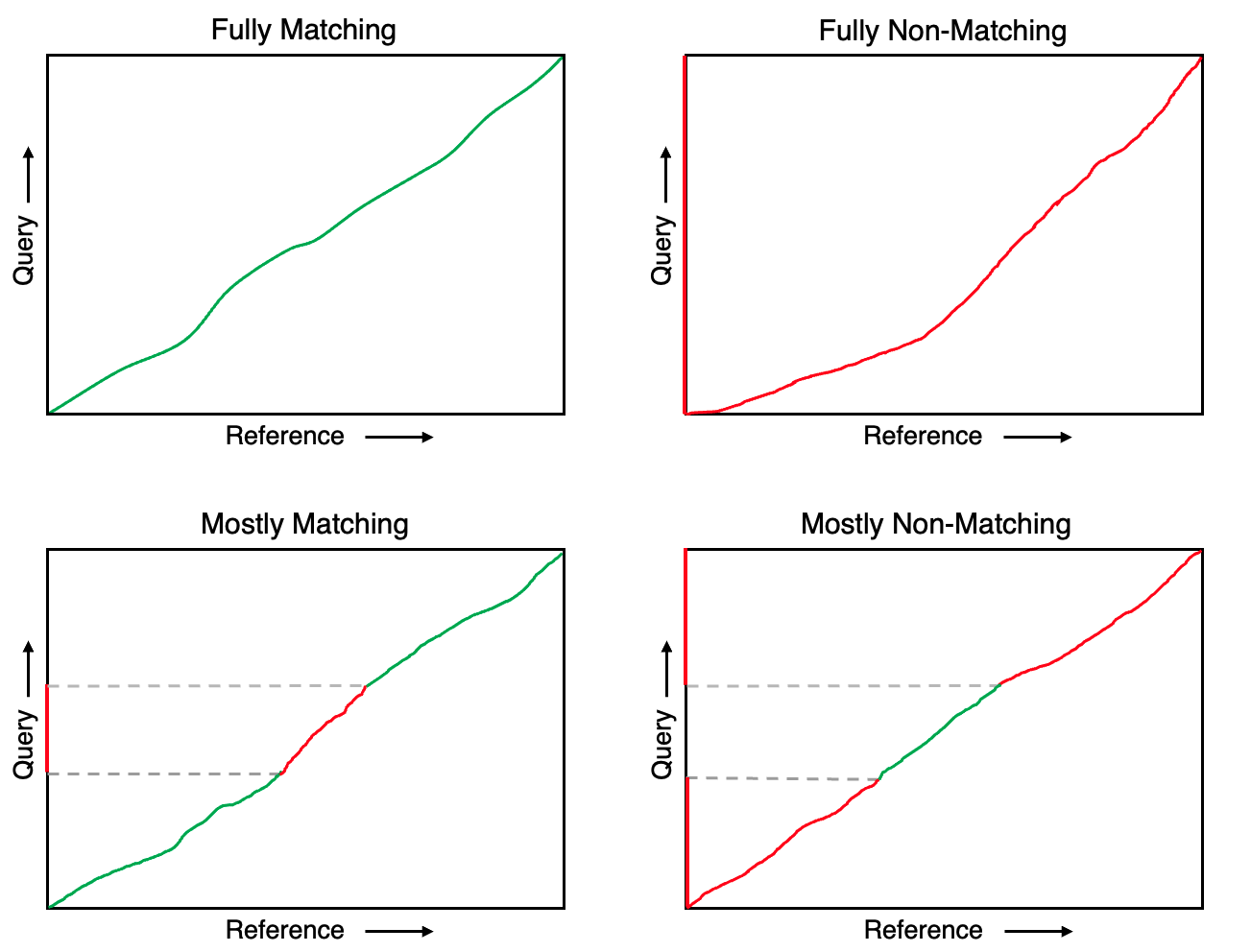}
 \caption{Overview of different scenarios in the benchmark suite.  Portions of the query recording that are tampered are shown in red along the vertical axis.  The goal of the reliability metric is to accurately classify parts of the DTW alignment path as reliable or unreliable.}
 \label{fig:benchmarkScenarios}
\end{figure}

\textit{Modified Benchmarks:~Details}.  We construct the 19 benchmarks in the following manner. First, we randomly select $N$ recordings from each Mazurka and consider all $\binom{N}{2}$ pairings of the same Mazurka ($N = 63$ in training, $N = 34$ in test). Selecting $N$ recordings ensures that Mazurkas with a lot of recordings are not severely overrepresented in aggregate metrics of performance. For each pair of recordings, we consider one recording to be the query and one recording to be the reference. The reference recording is always kept unmodified in its original form. The query recording is modified in the following ways to introduce non-matching regions into the cost matrix. Figure~\ref{fig:benchmarkScenarios} shows a graphical depiction of the different benchmark scenarios.
\begin{itemize}
  \item Benchmark \#1: Fully matching/non-matching. In this benchmark, half of the alignment pairs are completely matching and the other half are completely non-matching. For non-matching pairs, we replace the query recording with a recording of a different Mazurka. Since some Mazurkas are much longer than others and might create a cost matrix that has no viable DTW path, we force non-matching recordings to be the same duration as the reference, either by truncating or repeating.  Note that performing DTW with allowable transitions $(1,1), (1,2), (2,1)$ imposes a maximum time warping factor of 2, so there will be no viable DTW path if the sequence lengths differ by more than a factor of 2.
  
  \item Benchmarks \#2-10: Mostly matching. In these benchmarks, we replace a segment of the query recording with a segment (of the same duration) from a different Mazurka. Thus, most of the pairwise cost matrix will have a valid warping path, but there will be a section that is non-matching. We consider three different lengths for the duration of tampering: 10\%, 20\%, and 30\% of the length of the query sequence. Likewise, we consider three different locations where the tampering occurs: at the beginning, middle, and end of the query recording. Note that these scenarios might arise in practice as a result of silence/applause at the beginning or end of a recording, extended silence between movements, a different cadenza, etc. With three different durations and three different locations, this produces a total of 9 different benchmarks. Each benchmark has the same $2 \binom{63}{2} = 3906$ total pairs in training and $3 \binom{34}{2} = 1683$ total pairs in testing. The bottom left part of Figure 3 shows an example scenario from a mostly matching benchmark, where matching and non-matching regions are shown in green and red, respectively.

  \item Benchmarks \#11-19: Mostly non-matching. In these benchmarks, the pairwise cost matrix has a short matching region, and the rest of the matrix is non-matching. The query is constructed by first selecting a different Mazurka recording that is adjusted to match the duration of the reference recording. Then we replace a section of the query with a matching region. Again, we consider splicing durations of 10\%, 20\%, and 30\% of the query length, and we consider locations at the beginning, middle, and end of the query. With three durations and three locations, this produces a total of 9 different benchmarks. The bottom right part of Figure 3 shows an example scenario from a mostly non-matching benchmark.
  
\end{itemize}

\begin{table*}
	\centering
	\begin{tabular}{|l|ccc|ccc|}
		\hline
		\multirow{2}{*}{Benchmark} & \multicolumn{3}{c|}{Proposed} & \multicolumn{3}{c|}{Baseline}\\
		& AUROC ($\uparrow$) & TPR@2\%FA ($\uparrow$) & EER ($\downarrow$) & AUROC ($\uparrow$) & TPR@2\%FA ($\uparrow$) & EER ($\downarrow$) \\
		\hline
		Full match/non-match & .978 & 94.1\% & 4.1\% & .929 & 31.9\% & 14.0\% \\
		\hline
		Mostly match 10\% & .945 & 40.3\% & 8.9\% & .869 & 8.6\% & 20.2\% \\
		Mostly match 20\% & .958 & 69.6\% & 6.7\% & .897 & 14.1\% & 17.3\% \\
		Mostly match 30\% & .965 & 83.2\% & 5.9\% & .917 & 22.1\% & 15.0\% \\
		\hline
		Mostly non-match 10\% & .953 & 76.0\% & 7.8\% & .925 & 36.1\% & 14.4\% \\
		Mostly non-match 20\% & .964 & 88.3\% & 6.4\% & .932 & 36.0\% & 13.6\% \\
		Mostly non-match 30\% & .966 & 89.7\% & 6.0\% & .933 & 35.7\% & 13.5\% \\
		\hline
	\end{tabular}
	\caption{Performance of the proposed reliability metric ($L=300$, $\Delta=\frac{L}{5}$) on classifying elements in the DTW alignment path as reliable vs unreliable.  The rows correspond to different benchmark scenarios.}
	\label{tab:mainResults}
\end{table*}

\textit{Evaluation}.  The evaluation requires careful consideration.  Though it may seem like the most straightforward task, our goal is not to detect whether each part of the query is matching or non-matching.  Note that the DTW alignment path could be completely wrong in a matching region, but the reliability metric could still do its job well by correctly recognizing that the predicted alignment is unreliable.  In this case, DTW is performing poorly but the reliability metric performs well.  In our study, we are evaluating the performance of the reliability metric, not the performance of the DTW algorithm.

\textit{Evaluation Points: Motivation}.  Consider a DTW alignment path with reliability metric scores given by $(i_0, j_0, s_0)$, $(i_1, j_1, s_1)$, \dots, $(i_{T-1}, j_{T-1}, s_{T-1})$, where each coordinate $(i,j)$ has an associated reliability metric score $s$.  To evaluate the reliability metric score $s$, we must know if we are in a tampered region and, if not, how much alignment error there is.  Assume that the ground truth beat annotations for the unmodified query and reference are given by $(q_0, q_1, \dots, q_{K-1})$ and $(r_0, r_1, \dots, r_{K-1})$, respectively, where $K$ is the number of beats in the piece.  Before modifying the query, the ground truth alignments are given by the coordinates $(q_0, r_0)$, $(q_1, r_1)$, \dots, $(q_{K-1}, r_{K-1})$.  After modifying the query, the coordinates whose $q_k$ falls in a tampered region are discarded.  Let the remaining ground truth alignment points be denoted $(q_k,r_k)$ for $k \in \mathbb{A}$, where $\mathbb{A}$ contains the indices of ground truth coordinates that remain after the query is tampered.  Since we only know the ground truth alignment at selected points, we only evaluate the reliability metric scores at selected points where the alignment error is known.  

\textit{Evaluation Points: Procedure}.  Given the DTW alignment path with reliability metric scores $(i_0, j_0,s_0)$,\dots, $(i_{T-1},j_{T-1}, s_{T-1})$, we evaluate only a subset of these coordinates for which we can calculate accurate alignment errors.  This subset is selected in the following manner.  For each ground truth coordinate $(q_k,r_k)$, $k \in \mathbb{A}$, we determine the element on the DTW alignment path that is closest to $(q_k,r_k)$ in Euclidean distance.  The closest element is selected as an evaluation point, and the Euclidean distance indicates the alignment error.  Since this only selects elements in matching regions, we also uniformly sample elements on the DTW path in non-matching regions every $0.25$ seconds (which was selected to roughly match the separation between beat annotations).  Thus, in matching regions we select elements on the DTW path that are closest to each ground truth coordinate, and in non-matching regions we uniformly sample elements on the DTW path.

\textit{Ground Truth Labels}.  Consider an element $(i,j)$ on the DTW alignment path that is selected as an evaluation point.  The ground truth reliability label for the path element $(i,j)$ is determined in the following manner.  If the coordinate $(i,j)$ is in a non-matching region (i.e., frame $i$ is in a portion of the query that has been tampered), then the ground truth label is ``unreliable.''  If the coordinate $(i,j)$ is in a matching region (i.e., frame $i$ is in a portion of the query that has not been tampered), then the ground truth reliability label depends on how accurate the predicted alignment is.  If the alignment error at $(i,j)$ is less than $\varepsilon$ (the hyperparameter that defines the level of desired precision, described in the last paragraph of Section~\ref{sec:method}), then the ground truth reliability label is ``reliable.''  If the predicted alignment at $(i,j)$ is greater than $\varepsilon$, the ground truth reliability label is ``unreliable.''

\textit{Classification}.  We evaluate the reliability metric scores as a standard binary classification task.  We apply a threshold $\gamma$ to the reliability score to predict if that element in the warping path is reliable or unreliable.  For a fixed value of $\gamma$, we measure the true positive rate (TPR) and false positive rate (FPR) across all beats in all pairs of aligned recordings in the benchmark.  Here, the true positive rate indicates the fraction of reliable alignment coordinates that are correctly classified as reliable, and false positive rate indicates the fraction of unreliable alignment coordinates that are classified as reliable.  By sweeping across a range of $\gamma$ values, we can characterize the performance in a receiver operating characteristic (ROC) curve, which shows the tradeoff between the true positive rate and false positive rate.  Note that ROC curves tend to be used when classes are roughly balanced, whereas precision-recall curves are used when positives are rare.  Since our class balance ranges from roughly $0.1$ to $0.9$ (depending on the benchmark) and both classes are treated symmetrically across the benchmark suite, we characterize performance through an ROC curve.

\textit{Metrics}.  From the ROC curve, we extract several metrics of performance:
\begin{itemize}
  \item EER: The equal error rate indicates the point on the ROC curve where the false alarm rate and missed detection rate (MDR = 1 - TPR) are equal. It has the benefit of being invariant to class priors.
  \item AUROC: The area under the ROC curve measures overall discrimination ability and ranges between 0.5 (random) and 1.0 (perfect).
  \item TPR when FPR$=2\%$: TPR indicates the fraction of matching beats that are correctly classified. This is of interest to us in applications where we have weakly labeled data (e.g., different performances of the same piece) and would like to identify reliable alignments that can be used for training data. This metric tells us how much of the reliable alignments we utilize, while ensuring a high level of quality in the data. 
\end{itemize}

\section{Results}
\label{sec:results}

Table~\ref{tab:mainResults} shows the detection performance of the proposed reliability metric on the benchmark suite.  The rows correspond to different benchmark scenarios, where the ``Mostly match $X\%$'' and ``Mostly non-match $X\%$'' rows are averages of the beginning/middle/end benchmarks.  The left panel shows the AUROC, TPR@2\%FPR, and EER of the proposed reliability metric with $L=300$ and $\Delta = \frac{L}{5}$.  (We present experimental results justifying this choice of $L$ and $\Delta$ in Section~\ref{sec:analysis}.)  The right panel shows the performance of a naive baseline which simply uses the pairwise cost as a reliability metric.  Specifically, for each coordinate $(i,j)$ on the DTW alignment path, we simply use $-C[i,j]$ as an indicator of reliability, where the negative sign ensures that low pairwise costs indicate higher reliability.

There are three things to notice about Table~\ref{tab:mainResults}.  First, the proposed reliability metric performs much better than the baseline.  For example, on the fully matching/non-matching benchmark, the proposed reliability metric decreases EER from $14.0\%$ to $4.1\%$ and improves TPR at $2\%$ FPR from $31.9\%$ to $94.1\%$.  Second, the reliability metric is not extremely accurate but relatively consistent and good enough to be useful.  Across all of the scenarios, the metric achieves EERs between $4 - 9\%$ and AUROC scores between $0.94 - 0.98$.  Third, the reliability metric struggles to correctly identify matching or non-matching regions that are short in duration.  We see that the performance improves substantially as we progress from the $10\%$ mostly match benchmark to the $20\%$ and $30\%$ mostly match benchmarks, and likewise with the mostly non-match benchmarks.

\begin{figure}
	\centering
	\includegraphics[alt={A bar graphs showing Recall and AUROC increasing as window length L grows from 100 to 300 and then falling as L goes to 900, with the smallest hop size (L/5) consistently achieving the highest performance across all window sizes.},width=\linewidth]{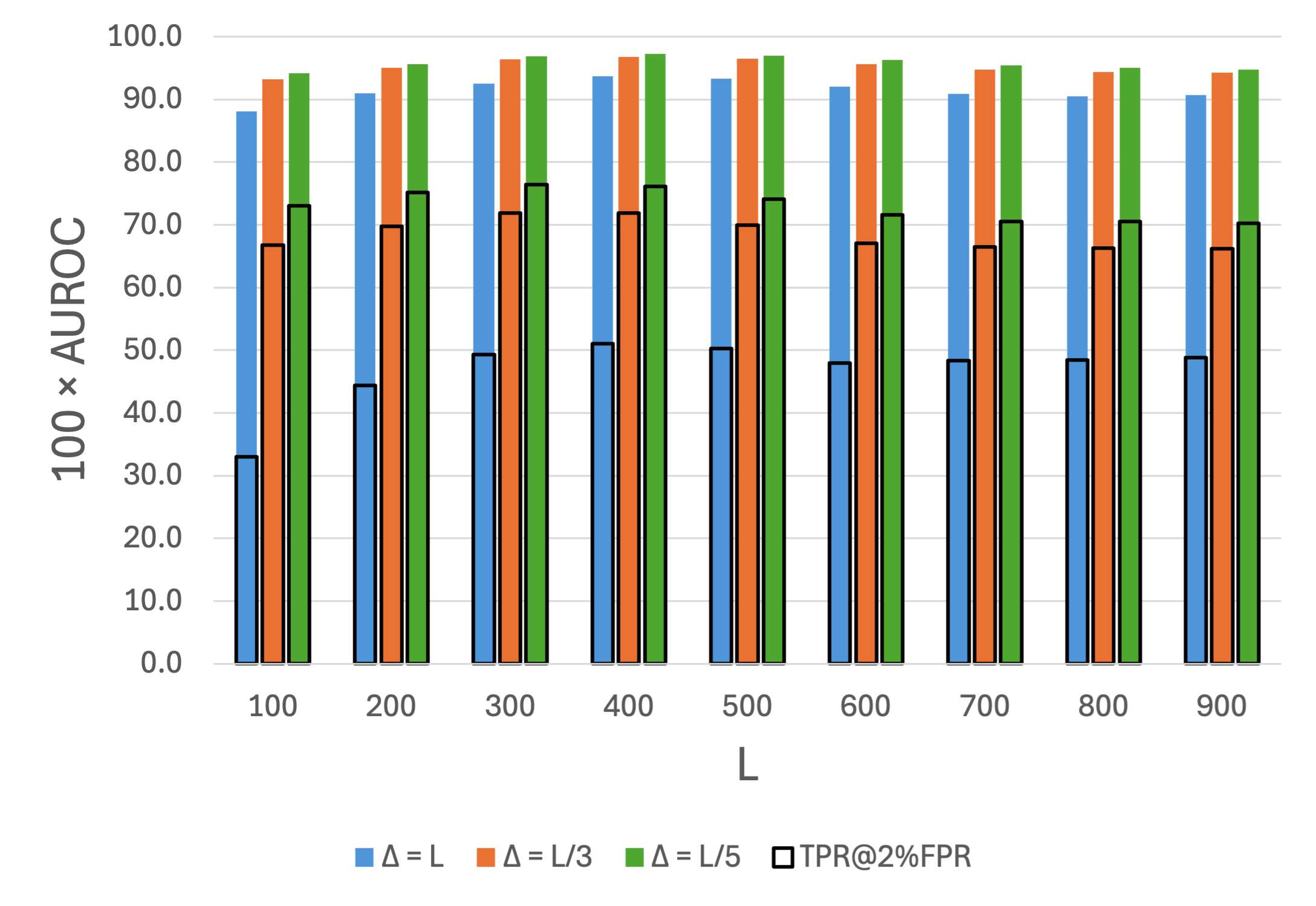}
	\caption{Characterizing the impact of chunk length $L$ and hop size $\Delta$ on the performance of the reliability metric.  Bar heights show TPR at $2\%$ FPR (black border) and $100 \times$ AUROC (solid fill) on the training/development data.}
	\label{fig:selectingHyperparameters}
\end{figure}

\section{Analysis}
\label{sec:analysis}

In this section we present two analyses to develop deeper intuition about the proposed reliability metric.

The first analysis is to characterize the effect of chunk size $L$ and hop size $\Delta$ on the performance of the reliability metric.  Figure~\ref{fig:selectingHyperparameters} compares the performance of the reliability metric on the training/development benchmark suite for several settings for $L$ and $\Delta$.  Each bar indicates the TPR at 2\% FPR (black border) and AUROC (solid fill) for a selected setting.  To compute a single aggregate score across the entire benchmark suite, we first compute scores on each of the 19 benchmarks and then average the individual benchmark scores.  These results thus reflect the performance of the reliability metric across a wide range of scenarios.

There are two takeaways from Figure~\ref{fig:selectingHyperparameters}.  First, a smaller hop size leads to better performance at the cost of increased computation.  We see a big improvement in performance when going from $\Delta=L$ to $\frac{L}{3}$, and a marginal improvement from $\Delta=\frac{L}{3}$ to $\frac{L}{5}$.  Note that $\Delta=L$ means that chunks are not overlapping, there is no ``voting'' among overlapping chunks, and the resolution in time is very poor.  When $\Delta=\frac{L}{3}$, most elements in the DTW path receive three ``votes'' from overlapping chunks, and the resolution in time is improved by a factor of $3$.  Second, a bigger chunk length leads to better performance up to $L=300$, but then performance degrades slightly as $L$ increases beyond that.  This is because a bigger chunk length has poorer resolution in time, which leads to misclassifications on matching or non-matching regions that are shorter in duration.  Based on these experiments, we selected $L=300$ and $\Delta=\frac{L}{5}$ for our test benchmark simulations in Table~\ref{tab:mainResults}.


The second analysis is to understand failure modes for the reliability metric.  One failure mode is when a chunk has multiple strong alignment paths and selects a different path than DTW.  For example, given a chunk in a matching region, if the first 3 seconds of the chunk match the last 3 seconds, there will be strong alignment paths in the upper left, bottom left, and bottom right corners of the cost matrix (assuming the same axis directions as in Figure~\ref{fig:methodOverview}).  Thus, homogeneous regions or musical repetition can lead to completely incorrect reliability scores.  Another failure mode is misclassifying elements close to a tampering boundary.  Because we impute the agreement metric to all DTW coordinates in the chunk, our method has relatively poor time resolution.  Our recommended setting of $L=300$ corresponds to a chunk length of approximately $7$ seconds, so the method will struggle to correctly identify short segments of matching or non-matching data with durations that are a few seconds or less.

\section{Conclusion}

We have proposed a metric that indicates the reliability of DTW alignment paths. Our reliability metric is based on circumstantial evidence: if the DTW alignment path is prominent, then it should not change even if given more freedom in its boundary conditions. Thus, by measuring how much the alignment path changes, we can estimate its reliability. We measure the utility of the proposed metric on a suite of carefully constructed benchmarks where alignment paths contain both matching and non-matching regions. We find that the reliability metric is able to classify regions of DTW alignment paths as reliable or unreliable with an aggregate AUROC of $0.970$. Based on our empirical results, we propose a set of recommended settings for use with audio-audio alignment tasks. In future work, we plan to explore ways to improve how well the metric can detect matching or non-matching regions of short duration, and to explore how well the metric works with different kinds of music data.

\section{Acknowledgments}
This material is based upon work supported by the National Science Foundation under Grant No.~2144050.

\bibliography{reliabilityMetric_ismir2026}

%
%
%
%

\end{document}